\documentclass{article}
\usepackage{graphicx}
\usepackage{amssymb}
\usepackage[utf8]{inputenc}

\usepackage[T1]{fontenc}
\usepackage{cite}

\def\arcsinh{\mathop{\mbox{arcsinh}}}

\begin{document}
\title{Quantum Gravity Wormholes\\ and Topological Teleporter}
\author{Igor Nikitin\\
Fraunhofer Institute for Algorithms and Scientific Computing\\
Schloss Birlinghoven, 53757 Sankt Augustin, Germany\\
\\
igor.nikitin@scai.fraunhofer.de
}
\date{}
\maketitle

\begin{abstract}
The question of a possibility of opening a wormhole due to the deformation of the equation of state of the matter caused by quantum gravity effects is considered. As a wormhole environment, the previously considered model of a galaxy with radial flows of dark matter is selected. The calculation shows, that for a wormhole solution in the framework of this model, after the nominal classical density exceeds a critical value in the Planck range, the components of the energy-momentum tensor become negative in a certain order. First, the mass density becomes negative, then the radial pressure, then the tangential pressure. In this case, at some point the flare-out conditions necessary and sufficient to open the wormhole are fulfilled. Calculations are made to open a wormhole in a model with parameters of the Milky Way galaxy. It is shown that the opening of the wormhole is accompanied by a separation of the bubble of space, completely detached from the wormhole and the outer space. In addition to this solution, a topologically dual one is found, looking like an instant swapping of two volumes in space, which can be interpreted as an event of teleportation.
\end{abstract}

\section{Introduction}\label{sec1}

Corrections of Quantum Gravity (QG) to Friedmann-Lemaître-Robertson-Walker (FLRW) cosmological model with a scalar field were studied by Ashtekar et al. in \cite{0602086,0604013,0607039}. The corrections lead to the modification of the mass density function participating in the equations:
\begin{eqnarray}
&&\rho=\rho_{nom}(1-\rho_{nom}/\rho_{crit}),\label{rhoqg}
\end{eqnarray}
where $ \rho_{nom} $ is the nominal classical density before the corrections, $ \rho_{crit} $ is the critical density value of the Planck order, $ \rho_{crit} \sim \rho_P $. According to this formula, when the nominal density exceeds the critical value, $ \rho_{nom}> \rho_{crit} $, the effective density becomes negative, $ \rho <0 $. In Rovelli and Vidotto \cite{1401.6562} and Barceló et al. \cite{1409.1501}, this result was used to construct the model of {\it Planck stars}, where, in the gravitational collapse of stars, the effect of gravitational repulsion arises, when the nominal density exceeds the critical value. This effect leads to the {\it quantum bounce} phenomenon: the collapse process reverses in time, all collapsing material is thrown out, the black hole becomes white.

To be strict, we note that in \cite{0602086,0604013,0607039} in the framework of QG corrected FLRW model, the Friedmann equation has the form $ H^2=8 \pi \rho / 3 $, where $ H $ is the Hubble parameter. This does not allow negative densities and a bounce occurs at $ \rho = 0 $, $ \rho_{nom} = \rho_{crit} $. However, in other models, in particular for the Planck stars \cite{1401.6562,1409.1501}, this does not prohibit the density from going into the negative region along the same or a similar curve, leading to the appearance of a gravitational repulsive force, which drives the quantum bounce effect.

\begin{figure}
\begin{center}
\includegraphics[width=0.6\textwidth]{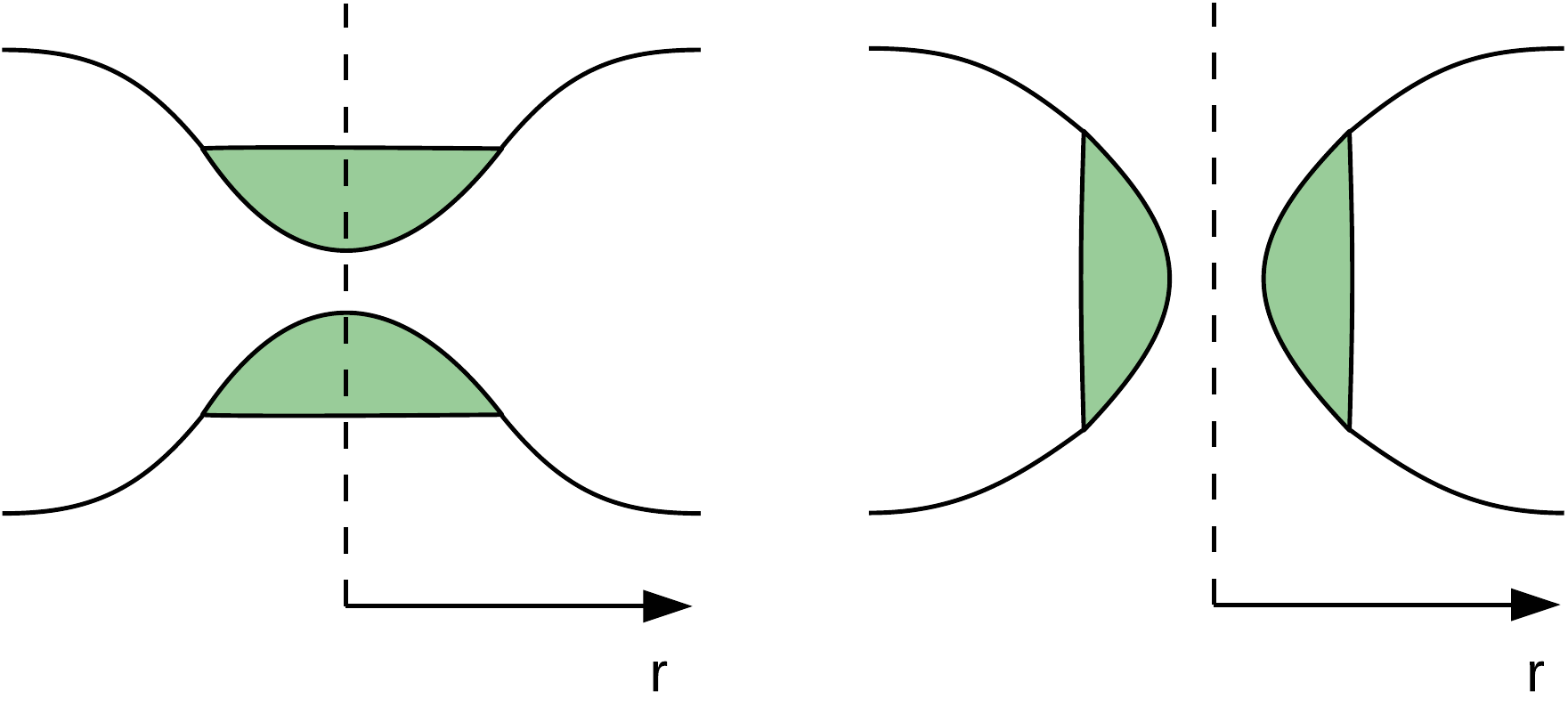}

\end{center}
\caption{Principal scheme of wormhole opening.}\label{f6}
\end{figure}

\begin{figure}
\centering
\includegraphics[width=0.8\textwidth]{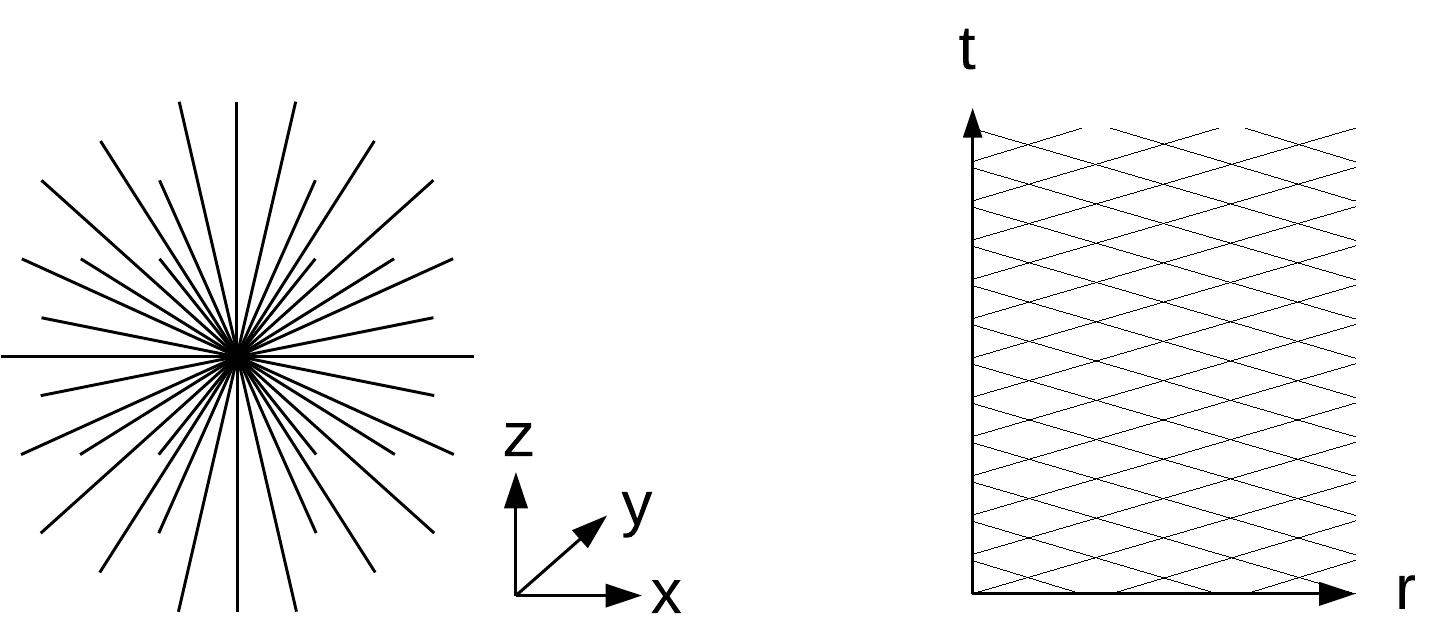}
\caption{RDM-star: a compact massive object, coupled to radially directed flows of dark matter. Image from~\cite{1701.01569}. Used here as an environment of the wormhole.}
\label{f0}
\end{figure}

The appearance of effectively negative density in the cores of collapsing stars means formation of {\it exotic matter}. It is widely known that {\it wormholes} also require exotic matter for their existence. In this paper, we will investigate the question of whether a quantum modification of the equations similar to (\ref{rhoqg}) can lead to the opening of a wormhole. In general form, this process is shown in Fig.\ref{f6}. Initially, the exotic matter (or QG effect, implementing it) is localized in the minimum of the gravitational potential. After the formation of a wormhole the exotic matter is localized in the throat, where the minimum of gravitational potential is also located, for the symmetric solutions under consideration. Thus, the exotic matter is localized exactly where it is needed to open and support the wormhole. We will see further that such a general scheme implements itself in our model, with certain differences in the topological details of the rearrangements.

As a wormhole environment, we will use the model of radial dark matter (RDM), constructed in our paper \cite{1701.01569}. This model considers the static configuration shown in Fig.\ref{f0}. It consists of radially directed energetically balanced incoming and outgoing flows of dark matter. It was shown in \cite{1701.01569} that a model of a galaxy with such a distribution of dark matter has flat rotation curves. Other aspects of this model were investigated in \cite{1707.02764,1811.03368,1812.11801,1903.09972,1906.09074}.

In this paper, in Section~\ref{sec2} we construct a wormhole, which, at the outer radius, where the nominal density reaches Planck values, is connected with the RDM model. We calculate the energy-momentum tensor, its density and pressure components $ (\rho, p_r, p_t) $, and obtain the equation of state (EOS) that relates these quantities to $ \rho_{nom} $. We show that the {\it flare-out conditions} necessary and sufficient to open a wormhole are naturally consistent with QG deformed EOS of type (\ref{rhoqg}). In Section~\ref{sec3} we will consider a realistic example, matching the model parameters with the sizes of the Milky Way galaxy.

In Section~\ref{sec4} we will consider the topological aspects of the wormhole opening process. We show that under certain natural assumptions, such as the finiteness of the energy-momentum tensor, a bubble of space (a baby universe), disconnected from the wormhole and outer space, arises in the process with necessity. We will also consider a topologically dual process that is equivalent to instant swapping of two volumes in space and that can be interpreted as {\it teleportation}. Unlike the well-known quantum teleportation, this process is entirely described by classical General Relativity. This configuration is similar to a wormhole, but differs by the absence of a throat and by the type of matter distribution. Since this process is based on topological restructuring of space, reconnection of maps, we designate it as {\it topological teleportation}.

\section{QG wormhole}\label{sec2}

To describe the solutions, the standard metric of a static spherically symmetric type is chosen:
\begin{equation}
ds^2=-A(r)dt^2+B(r)dr^2+r^2(d\theta^2+\sin^2\theta\; d\phi^2).\label{stdmetr}
\end{equation}

Here $ t $ represents the time of the remote observer, $ r $ is aerial radius. A sphere of radius $ r $ has an area of $ 4 \pi r^2 $ by definition. $ (\theta, \phi) $ represent the angles of the spherical coordinate system. Two profiles $ A (r) $, $ B (r) $ define the spacetime deformation. The $ A $ profile defines the time dilation and the associated effects of gravitational redshift. The $ B $ profile defines the non-Euclidean radial deformation of space, namely, the deviation from $ r $ of the radial proper length integral $ L = \int_0^r \sqrt {B} dr $. The positivity of the profiles $ A> 0 $, $ B> 0 $ is equivalent to the absence of trapping horizons in the solution.

In this paper, we will explore the possibility of opening a wormhole in the RDM model. This model is characterized by the absence of a transversal pressure component $ p_t = 0 $ and the presence of extremely high densities and radial pressure components in the center of the system:
\begin{eqnarray}
&&\rho=p_r=\epsilon/(8\pi r^2A).\label{rhopeff}
\end{eqnarray}

Here, for definiteness, we considered a null type of radial dark matter (NRDM), which is equivalent to a perfect fluid with EOS of the form $ \rho = p_r $, $ p_t = 0 $. The constant factor $ \epsilon> 0 $ determines the overall scaling of the solution. Redshift factor $ A $ in the central region of the RDM solution has an extremely rapid drop with decreasing $ r $, which was denoted in \cite{1701.01569} as {\it red supershift}. This behavior, together with the $ r^{-2} $ factor, leads to extremely high values of the classical energy density at the center of the system. This makes this system a potential candidate for QG corrections, which can effectively lead to negative densities and pressures in the Einstein field equations (EFE) and, ultimately, to the opening of a wormhole.

The $ B (r) $ profile plays an important role in analyzing wormhole solutions. It is associated with the Misner-Sharp mass function (MSM):
\begin{eqnarray}
&&M=r/2\ (1-B^{-1}).
\end{eqnarray}

A point with a minimum radius value, {\it a throat} of the wormhole has a finite positive value $ A $ and an infinite positive value $ B $, which corresponds to
\begin{equation}
A(r_0)>0,\ B(r_0)\to+\infty,\ 2M(r_0)=r_0.\label{r0def}
\end{equation}

Note that for solutions with trapping horizons also $ B \to + \infty $ for $ r \to r_0 + 0 $, but when passing through the point $ r_0 $, the sign of the profiles changes simultaneously, $ A <0 $ and $ B < 0 $. For wormhole solutions, $ A (r_0)> 0 $ is fixed. Moreover, only one profile $ B (r) $ cannot change sign, since this would change the signature of the metric from $ (- +++) $ to $ (- - ++) $, which we do not allow in the considered class of solutions. As a result, the solution cannot be continued to the left of $ r_0 $, this value of the radius is really minimal. The solution can be stitched at the point $ r_0 $ with another, similarly behaving solution, in the particular case with its copy. Moreover, the irregularity of the solution in the throat can be corrected by a change of coordinates $ r \to L $, where the proper length integral $ L = \pm \int_0^r \sqrt {B} dr $ is finite on the solutions under consideration, the sign is chosen differently for different copies and the integration constant is chosen to provide $ L (r_0) = 0 $. After such a replacement, the function $ r (L) $ turns out to be smooth and even, as well as $ A (r (L)) $, $ \rho (r (L)) $ and other functions depending on it. This method defines a symmetrical continuation of the wormhole, described by Visser in \cite{Visser1996}. Although such a continuation is not unique, it is the most demonstrative and simply formulated one. It is also convenient for building wormholes that act within the same universe. In general, wormholes can connect different universes, which may even have different time rates for the observers distant from the mass clusters, \cite{Visser1996}. In this paper, we will consider only the symmetric version of wormholes.

For calculations, we need a number of coordinate substitutions. This is necessary to regularize the behavior of profiles in the throat, as well as to display very large and very small values of the profiles $ A, B $ and radius $ r $ on the graphs. In work \cite{1701.01569} we managed this with logarithmic substitutions
\begin{eqnarray}
&&a=\log A,\ b=\log B,\ x=\log r,\label{abx}
\end{eqnarray}
\cite{1707.02764} introduced
\begin{eqnarray}
&&Z=B^{-1},\ W=\pm\sqrt{Z},
\end{eqnarray}
where $ Z $ transforms the infinite value of the profile $ B $ in the throat to zero, $ B \to \infty $, $ Z \to0 $, and different signs of $ W $ mark different copies of spacetime connected by the wormhole. Instead of the logarithmic transformation, the following scaling functions are used:
\begin{eqnarray}
&&f(h)=\arcsinh(h/2),\ f^{-1}(h)=2\sinh(h),\\
&&z=f(Z),\ w=f(W),\ y=f(r),\ l=f(L),
\end{eqnarray}
which allows to control the neighborhood of $ Z = 0 $, $ r = 0 $ points, makes transitions between $ \pm W $ copies and also possesses the logarithmic asymptotics $ f (v) \sim \log v $ at infinity.

The basic formulas are taken from Visser \cite{Visser1996}, Lobo \cite{1604.02082}, also compatible with the author's work \cite{1707.02764}:
\begin{eqnarray}
&&\rho=M'_r/(4 \pi r^2), \label{rpMA1}\\
&&p_r=(-2 A M + r (r - 2 M) A'_r)/(8 \pi r^3 A), \\
&&p_t=(-r^2 (r - 2 M) (A'_r)^2 + 4 A^2 (M - r M'_r) \\
&&+ 2 r A (-A'_r (M + r (-1 + M'_r)) + r (r - 2 M) A''_{rr}))/(32 \pi r^3 A^2),\label{rpMA2}
\end{eqnarray}
where for the Morris-Thorne (MT) shape function, we substituted its expression through MSM, $ b_{MT} \to2M $, and also used our definition of the redshift factor, $ 2 \Phi \to \log A $. The first formula here reflects the fact that the mass function is the integral over the radius of the product of the area by the energy density. The remaining formulas follow from EFE resolved for the components of the energy-momentum tensor, that is, the density and pressure values for the perfect fluid. There is also a relation linking these quantities, the hydrostatic equation
\begin{eqnarray}
&&p'_r=2(p_t-p_r)/r-(\rho+p_r)A'_r/(2A),
\end{eqnarray}
which implements energy-momentum conservation and is satisfied automatically for $ (\rho, p_r, p_t) $ given by the formulas (\ref{rpMA1})-(\ref{rpMA2}).

The definitions of $ (\rho, p_r, p_t) $ can also be re-expressed in terms of the functions $ (A, Z) $ and their derivatives
\begin{eqnarray}
&&\rho=-(-1 + Z + r Z'_r)/(8 \pi r^2), \label{rpAZ1}\\
&&p_r=(A (-1 + Z) + r Z A'_r)/(8 \pi r^2 A),\\
&&p_t=(-r Z (A'_r)^2 + 2 A^2 Z'_r \\
&&+ A (r A'_r Z'_r + 2 Z (A'_r + r A''_{rr})))/(32 \pi r A^2).\label{rpAZ2}
\end{eqnarray}

The so-called flare-out conditions for opening a wormhole \cite{Visser1996,1604.02082} directly follow from these equations. Here we will consider their special version, valid under the monotonicity condition $A'_r>0$, implemented in our model. Then, the following conditions hold in the throat: 
\begin{eqnarray}
&r=r_0,\ Z=0,\ Z'_r>0\ \Leftrightarrow\ p_r=-1/(8\pi r^2)<0,\ \rho+p_r<0\label{flareout}\\
&\Rightarrow\ p_t>0,\ \rho+p_r+2p_t>0.\label{flareout2}
\end{eqnarray}
The conditions in the first line here are necessary and sufficient for opening the wormhole, the conditions in the second line are the consequences of the first one (more necessary conditions). Here we used the definition (\ref{r0def}), as well as the increase in the function $ Z (r) $ near $ r_0 $. Indeed, $ Z (r) $ in the local vicinity of the throat $ r> r_0 $ increases, since its value changes from zero to positive. We also assume $ A'_r>0 $ everywhere. In addition, we see from these formulas that the values of $ \rho + p_r $ and $ p_t $ in the throat are proportional to $ Z'_r $ and can be made arbitrarily close to zero by choosing the appropriate $ Z $ profile, while $ p_r $ turns out to be negative with a margin. This margin depends only on the radius of throat and does not depend on the profiles.

We will also consider two more special cases for these formulas. One is realized at the bifurcation points considered below:
\begin{eqnarray}
&r=r^*,\ Z=0,\ Z'_r=0\ \Leftrightarrow\ p_r=-1/(8\pi r^2)<0,\ \rho+p_r=0\label{bif}\\
&\Rightarrow\ p_t=0,\ \rho+p_r+2p_t=0.\label{bif2}
\end{eqnarray}
the other is satisfied on the outer boundary of the spatial bubble:
\begin{eqnarray}
&r=r_0^*,\ Z=0,\ Z'_r<0\ \Leftrightarrow\ p_r=-1/(8\pi r^2)<0,\ \rho+p_r>0\label{bub}\\
&\Rightarrow\ p_t<0,\ \rho+p_r+2p_t<0.\label{bub2}
\end{eqnarray}

Further, we can reformulate the relations (\ref{rpAZ1})-(\ref{rpAZ2}) in terms of $ (a, z, y) $ and other functions introduced above, using chain differentiation rules. Thus, we are able to compute $ (\rho, p_r, p_t) $ for the metric given in the form $ (a (y), z (y)) $.

Our further strategy is as follows. We define the QG cutoff in the model under consideration as the point at which the nominal density given by the formula (\ref{rhopeff}) reaches a certain fraction of the Planck density:
\begin{eqnarray}
&&\rho_{nom}=\epsilon/(8\pi r^2A)= \rho_P/N. \label{QGcut}
\end{eqnarray}

Note that the formula for the nominal density is practically model independent, it encodes: the radial convergence factor $ r^{-2} $, the ultraviolet shift factor $ A^{- 1/2} $ for the radiation coming to the central region from large distances and another factor $ A^{- 1/2} $ for the acceleration of the proper time in the central region with respect to the time of the distant observer when calculating the energy power of such a radiation. The attenuation factor $ N \sim1-10 $ takes into account the possibility that the startup of QG effects may occur somewhat earlier than the achievement of the Planck density. In addition, in model examples, we can make these effects occur at moderately high densities.

After passing through the cutoff point, the QG effects modify the equations in a widely unknown way, which gives sufficient freedom for solution engineering. For example, in the Planck star model in \cite{1401.6562,1409.1501}, the interior of the QG region was engineered to continuously join the solutions describing black and white holes and reproduce the expected QG bounce effect. When constructing wormhole solutions, we will assume that QG effects modify EFE in such a way that only the EOS included in them changes. We will present this equation in a general form as the dependence of the actual components of the energy-momentum tensor included in the EFE on the nominal density determined above:
\begin{eqnarray}
&&\rho=\rho(\rho_{nom}),\ p_r=p_r (\rho_{nom}),\ p_t=p_t (\rho_{nom}).
\end{eqnarray}

Next, we will set the shape of the profiles $ (a (y), z (y)) $ to ensure the opening of the wormhole in the model. Then we inspect $ (\rho, p_r, p_t) $ profiles and find out whether they can be interpreted as QG deformed EOS, for which we will formulate a number of criteria. Thus, we will find a special solution for EOS, which may be the result of QG effects and leads to the opening of a wormhole. In fact, we consider general solutions, limiting them to the requirements to be static, spherically symmetric, and several technical requirements ($ \pm L $-symmetry, monotonicity and invariance of the $ A $-profile under deformations). Thus, there is wide functional arbitrariness in the found solutions, along with an explicitly constructed representative, a class of close QG wormhole solutions of the considered type will be found.

To be even more specific, we will explicitly build not only one wormhole solution, but also a solution sufficiently close to it without a wormhole, as well as solutions along the continuous path that connects them. For the obtained one-dimensional family of solutions, we formulate the criteria necessary for this family to describe the quasistatic wormhole opening process. Thus, we will come close to solving the question of the physical possibility of the dynamic process of opening a wormhole with a corresponding change of the topology of the space.

We proceed further as follows. First, we examine the fundamental possibility of opening a wormhole using a model example, not trying to reproduce the real physical scale of the system. Next, we repeat this calculation for a model with parameters of central black hole in the Milky Way galaxy and confirm that this process is possible for real physical scales.

\begin{figure}
\begin{center}
\includegraphics[width=\textwidth]{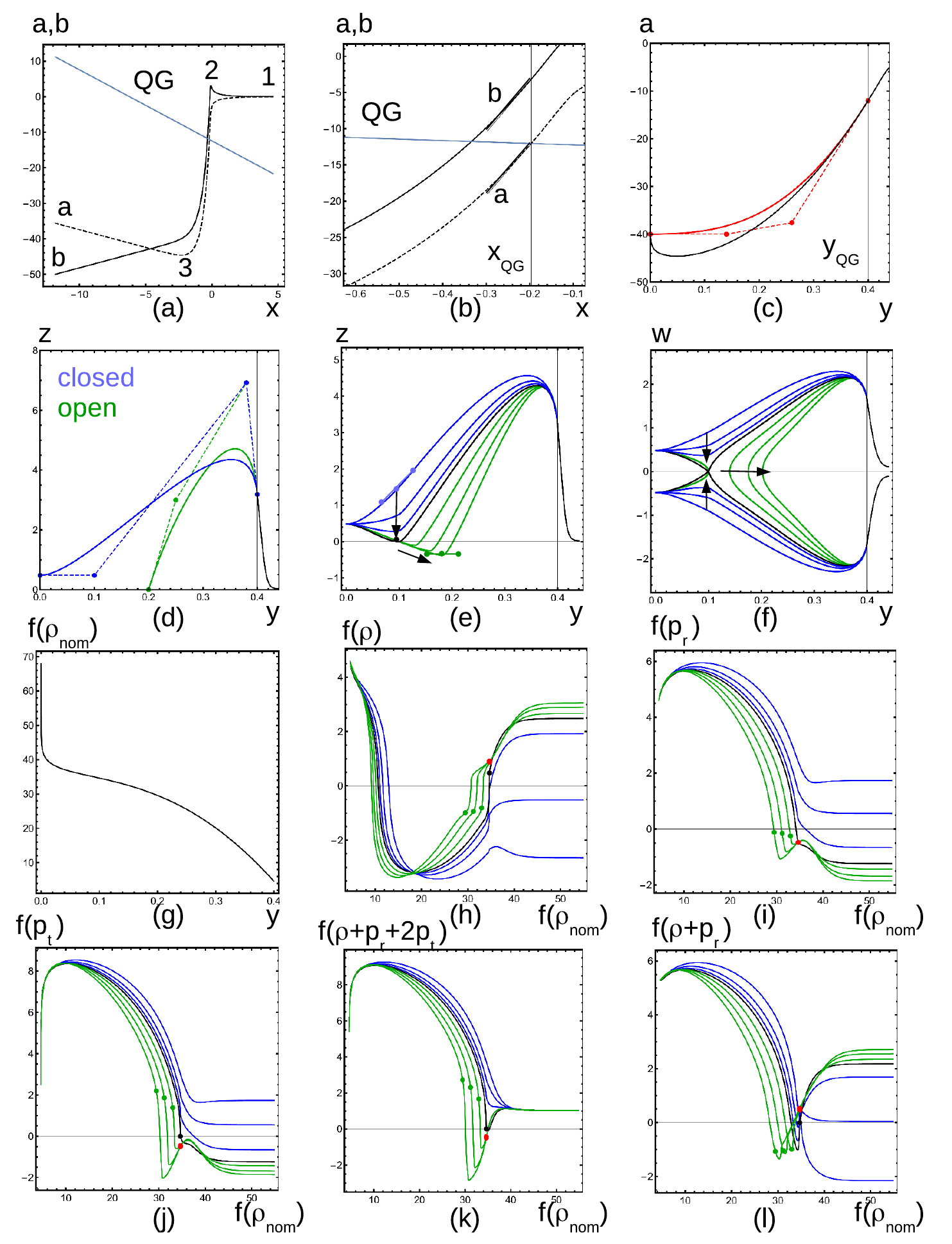}

\end{center}
\caption{Construction of QG wormhole (see text).}\label{f1}
\end{figure}

Fig.\ref{f1}a shows the NRDM solution with $ \epsilon = 0.01 $, $ r_s = 1 $. This solution was found using the numerical integration procedure described in \cite{1701.01569}. The $ r_s $ parameter determines the nominal gravitational radius of the system and is used when choosing a starting point. Integration starts from point 1 at a distance $ r_1 = 100 $, where the remote observer's clock $ a_1 = 0 $ and the value of $ b_1 $ corresponding to the gravitational radius of $ r_s $ are also set. Further, at point 2, the solution tries to enter the Schwarzschild mode, which is visible as raising $ b_2 $ and symmetrically falling $ a_2 $. Recall that the profiles are shown in a logarithmic scale and the metric coefficients $ A_2, B_2 $ at this point differ from the starting values by an order of magnitude. Then the supershift mode comes, in which both metric coefficients fall deeply, tens of orders of magnitude. This mode ends at point 3, at which a minimum of $ a_3 $ is reached. After that, both profiles diverge symmetrically, this part corresponds to the naked singularity of negative mass in the center of the system.

A physical interpretation of this solution can be obtained by considering the behavior of MSM. A rapid drop in the $ b $ profile corresponds to decreasing MSM (imagine that layers of positive mass are removed from the star). The value of the energy density is high positive. This behavior is similar to the previously investigated mass inflation phenomenon found in a charged black hole model with counterstreaming matter flows, by Hamilton and Pollack \cite{0411062}. In our case, a rapid decrease in MSM brings the solution to the region of negative masses. This means that a nucleus of negative mass is present in the center of the system. Since the energy density in the classical NRDM model remains positive all the time, the negative mass is concentrated in the naked singularity in the center of the system. On the other hand, we now impose the QG cutoff, which, starting from a certain radius, modifies the solution and regularizes the central singularity.

The QG cutoff is given by the formula (\ref{QGcut}) and is represented by the blue line in Fig.\ref{f1}a. For illustrative purposes, in this model example, we use a large attenuation factor in this formula, so $ \rho_P / N = 100 $ geometrized units. In Fig.\ref{f1}b the QG cutoff is shown in more detail, and the tangents that must be kept for $ C^1 $-stitching of profiles are also shown.

In Fig.\ref{f1}c, we are modeling $ a (y) $ profile. To define the profiles, we use Bézier curves, as described in Appendix. From the formulas (\ref{rpAZ1})-(\ref{rpAZ2}) the presence of the second derivatives of $ a $-profile in $ p_t $ is seen, therefore, $ C^1 $-stitching of $ a $-profile in the general case leads to a discontinuity of $ p_t $. We allow such discontinuity, interpreting it as a sharp startup of the transverse interaction between radially converging flows of dark matter that occur after the intersection of $ y_{QG} $. We also experimented with $ C^2 $-stitching of $ a $ -profile, which can be achieved by imposing the corresponding constraint on the Bézier curve. The result is a continuous but rapid increase in $ p_t $ from zero to a large positive value, which has the same physical interpretation. We conclude that there is physically no big difference when choosing the smoothness of the stitching of $ a $-profile among $ C^1 $ and $ C^2 $.

Next, by choosing the control points, we achieve that $ a (y) $ is a monotonically decreasing function with decreasing $ y $. From here and the formula (\ref{QGcut}), we see that $ \rho_{nom} (y) $ is monotonically growing with decreasing $ y $. Fig.\ref{f1}g shows this function using the scaling map $ f (\rho_{nom} (y)) $. The monotonicity property is convenient in our constructions, since the resulting dependences $ (\rho, p_r, p_t) (y) $ are not hard to transcode $ y \to \rho_{nom} $ and get the EOS in the desired form, as a function of $ \rho_{nom} $. We also note that $ a (y) $-profile can change during evolution, but for simplicity we consider a particular class of solutions with a fixed profile $ a (y) $.

Fig.\ref{f1}d shows $ z (y) $ profile for two wormhole states, the closed state is shown in blue, and the open state is shown in green. On the QG cutoff boundary, the profiles have a common tangent corresponding to $ C^1 $-continuation from the classical domain. For some time, the modified $ z (y) $-profile continues to grow rapidly with decreasing $ y $, which corresponds to a decreasing $ M (y) $ and positive density, as in NRDM. Further, the profile begins to decrease, which marks the boundaries of the regularized core of the effectively negative mass, required in the model for its self-consistency. At this point, the QG effects should do their job and make the density negative. After that, the blue and green profiles diverge.

The closed state is characterized by the reachability of the point $ y = r = 0 $. In addition, we require that the effective density be bounded in the vicinity of the point $ r = 0 $. As a result, near the center, $ M \sim4 / 3 \pi \rho (0) r^3 $ and $ Z = 1-2M / r \sim 1-8 / 3 \pi \rho (0) r^2 $, that is, $ z (y) $ for $ y \to0 $ tends to the constant $ z (0) = f (1/2) \approx0.481212 $ with a horizontal tangent. If we put $ \rho (0) <0 $, as it should be for the exotic core with completely negative density, then the profile will tend to this tangent from above, as shown in Fig.\ref{f1}d.

For the open state, the profile decreases directly to $ z \to + 0 $, which corresponds to the formation of a wormhole throat. The position at which this occurs corresponds to the freely adjustable parameter $ y_0 $.

Now we will ask how it is possible to make a continuous transition from the blue curve to the green one in Fig.\ref{f1}d, while maintaining the above boundary conditions. In the following sections, we will consider some alternatives. Now we will describe the transition, which is the most correct, in our opinion. It consists of two phases, shown in Fig.\ref{f1}e.

In the first phase, $ z (y) $-profile is bent, leading to the formation of a small core of positive density inside a large core of negative density. Then the profile reaches the minimum value $ z = 0 $ at the point $ y^* $. At this point, a bifurcation occurs in the model and the wormhole opens.

In the second phase, $ z (y) $-profile deepens into the negative region. Note that the physical part of the solution is localized at $ z \geq0 $. The control points of the Bézier curve used to set the profile go to the negative region, along with the corresponding continuation of the profile. At first glance, only $ z \geq0 $ values should be used in the formulas (\ref{rpAZ1})-(\ref{rpAZ2}) to describe the physical part of EOS. On the other hand, the resulting EOS will have a gap in place of $ z <0 $. This gap can be filled arbitrarily, with any values that continuously interpolate the gap and are not actually used. In particular, they can be filled with values calculated by the formulas (\ref{rpAZ1})-(\ref{rpAZ2}) for $ z <0 $. Since these formulas are nonsingular when passing through $ z = 0 $, this method gives one of the possible continuous interpolations of the gap.

Note that the diagram in Fig.\ref{f1}e describes an interesting topological rearrangement in which a wormhole forms suddenly at a nonzero radius, and the space inside this radius sticks together with its copy on the boundary sphere and forms a closed bubble that is completely disconnected from the external space. Matter and observers there will no longer be able to return to the outer space, except as a result of the reverse procedure of gluing the bubble to a suitable wormhole, performed in a sophisticated rescue operation.

We also note that after formation, the bubble can change its internal profiles, but cannot change its external radius. This is due to the fact that the total energy, that is, the MSM of this isolated system is conserved. By virtue of $ Z (r^*) = 1-2M (r^*) / r^* = 0 $ we get $ M (r^*) = r^* / 2 = Const $. The moment of detachment of the wormhole from the bubble is also interesting. According to \cite{Visser1996}, the wormhole throat is associated with its own positive mass, which is exactly $ M (r^*) = r^* / 2 $. This mass is not related to any density of matter, but arises as a constant contribution to MSM, due to $ Z (r^*) = 0 $. On the other hand, just before the opening of the wormhole, namely such mass was enclosed by $ r=r^* $ sphere. After the reconnection, the mass inside the sphere is preserved, while the wormhole throat radius can change and, for example, increase.

Technically, at the construction level of the Bézier curve for the second phase, we constrain its parameters so that it passes through the point $ y^* $ all the time. The obtained Bézier parameters represent a data set by which it is possible to reproduce the selected profiles unambiguously, and we present them in the Table~\ref{tab1} along with other characteristics of the model.

\begin{table}
\begin{center}
\caption{QG wormhole, model example}\label{tab1}

~

\def\arraystretch{1.1}
\begin{tabular}{|l|}
\hline
Model parameters: $\epsilon=0.01$, $r_s=1$
\\ \hline
QG cutoff: $\rho_P/N=100$, $r_{QG}=0.821337$, $x_{QG}=-0.196821$, \\
$y_{QG}=0.399923$, $a_{QG}=-12.0409$, $b_{QG}=-3.1829$, $z_{QG}=3.18461$, \\ 
$\{y_i,a_i\}=\{\{0.399923, -12.0409\}, \{0.259923, -37.5963\}, \{0.14, -40\}, \{0, -40\}\}$
\\ \hline
Closed state: $\{y_i,z_i\}=\{\{0.399923, 3.18461\}, \{0.379923, 6.92349\}, $ \\
$\{0.13, 1.95\}, \{0.1, 1.5\}, \{0.07, 1.05\}, \{0.03, 0.481212\}, \{0, 0.481212\}\}$, \\
redshift factor in the center $a(0)=-40$
\\ \hline
Open state: $\{y_i,z_i\}=\{\{0.399923, 3.18461\}, \{0.379923, 6.92349\}, $ \\
$\{0.212109, -0.38\}, \{0.182109, -0.38\}, \{0.152109, -0.38\}, $ \\
$\{0.03, 0.481212\}, \{0, 0.481212\}\}$, \\
redshift factor in the throat: $a(r_0)=-35.4799$,\\
radius of the throat: $r_0=0.406882$,\\
radius of the bubble: $r^*=0.200334$
\\ \hline
\end{tabular}

\end{center}
\end{table}

Fig.\ref{f1}f shows the profiles $ w (y) $, depicting the process of opening a wormhole. It is seen that at the time of opening, a hyperbolic rearrangement occurs, leading to reconnection of the patches of space and the formation of the wormhole and the bubble. Then the bubble inflates a little and approximately retains its shape, and the throat radius increases until it reaches the target value $ r_0 $ for the previously considered final open state.

In the following sections, we will consider in more detail the topological reconnection we have found, as well as possible alternatives. Now we only note that the methods we used are limited to the quasistatic case, but can be extended to non-stationary processes. Then, for the evolution of the bubble, other options can be implemented, such as its expansion or contraction in the framework of closed cosmological models.

Finally, Fig.\ref{f1}h,i,j show the dependences of the resulting $ (\rho, p_r, p_t) $ on $ \rho_{nom} $, which represent the corresponding EOS. In the figures, green lines indicate open states, blue lines indicate closed states, and black lines indicate the boundary corresponding to the bifurcation of the solution. Green points indicate the positions of the wormhole throat $ y_0 $, red points indicate the boundary of the bubble $ y_0^* $, and black points (which sometimes coincide with red ones) indicate the positions of the bifurcation point $ y^* $. From the position of the points in the figures, one can directly see the fulfillment of flare-out conditions (\ref{flareout}),(\ref{flareout2}), bifurcation conditions (\ref{bif}),(\ref{bif2}) and conditions at the bubble boundary (\ref{bub}),(\ref{bub2}). Recall also that between the red and the green points the space-time in our model has a gap, and the corresponding EOS curves in this segment can be omitted.

Note that in the approach we are considering, opening a wormhole is accompanied by a change of EOS. Physically, EOS may depend on temperature and other internal parameters, their change can lead to the necessary change of EOS. Another approach is possible, with fixing EOS and changing boundary conditions, for example, NRDM pressure on the QG boundary or the global parameter $ \epsilon $ controlling it. This approach requires the numerical integration of differential equations and is more difficult to implement. In the meantime, we are choosing a simpler engineering approach, we are investigating what form the resulting EOS has and, most importantly, whether it has common features with the previously explored QG deformed EOS.

On the left side of these graphs, when entering the QG region, $ \rho = p_r $ is fulfilled as a consequence of the continuity of stitching with the classical mode. The value of $ p_t $ undergoes a jump from zero to a positive value, which, as we have already discussed, is related to the choice of the smoothness class $ C^1 $ for stitching $ a $-profile. We interpret this as the startup of the tangential interaction between the radial flows of dark matter, occurring due to QG effects.

Further, after entering QG mode, $ \rho $ starts to fall and becomes negative. This behavior is characteristic of both the closed and open state and is associated, as we have already noted, with the necessary presence of the negative mass core inside the RDM model, in our case regularized due to QG effects. The departure of the density to the negative region is also consistent with the formula (\ref{rhoqg}) obtained in \cite{0602086,0604013,0607039}, as well as with the quantum bounce models \cite{1401.6562,1409.1501} using it.

Further on our curves, we see that in the three final open states, following $ \rho $, the value of $ p_r $ becomes negative. At the same time, $ p_t $ begins to decrease, but still remains positive in the area of interest. As a result, zones are formed in which the conditions (\ref{flareout}),(\ref{flareout2}) are met to open the wormhole.

Note that in the model of Planck stars \cite{1401.6562,1409.1501}, the argument about the gravitational repulsion force arising when the Planck density is exceeded imposes certain conditions on the pressure of the system, since the gravitational field is created not only by the density, but also by the pressure. For example, when analyzing Friedmann and Tolman-Oppenheimer-Volkoff equations, Blau \cite{Blau2018} noted that density enters into the equations in combinations with pressure components $ \rho + p_r $ and $ \rho + p_r + 2p_t $. We see that these combinations also play an important role in formulating flare-out conditions (\ref{flareout}),(\ref{flareout2}). The behavior of these combinations is shown in Fig.\ref{f1}k,l. We see that in zones of restructuring, $ \rho + p_r $ is stably negative and $ \rho + p_r + 2p_t $ is stably positive for open states and the transition occurs shortly after $ p_r $ goes into the negative region.

The right side of the graphs corresponds to the effect of separation of the bubble. We see how $ \rho $ near the center again becomes positive, which is necessary for opening a wormhole in the considered scenario. Thus, in the given example, in order to obtain the described process in the quasistatic mode, it is necessary that the QG deformed EOS does not follow simple law (\ref{rhoqg}), but after passing negative values changes its direction and moves to the positive region again. Note that only the bubble formation region requires such behavior, while the region near the wormhole has a stably negative $ \rho $.

To reproduce the necessary behavior of $ \rho $, the model has a lot of possibilities. First of all, the formula (\ref{rhoqg}) was obtained in \cite{0602086,0604013,0607039} when taking into account the quantum corrections in the leading order and, strictly speaking, in another model (cosmology of scalar fields). It is possible that higher-order calculations and a change of the model to NRDM can give the required behavior of the $ \rho $-profile. Another possibility is to fix the EOS of type (\ref{rhoqg}) and consider nonmonotonic profiles $ a (y) $ and $ \rho_{nom} (y) $. Such profiles can bring the solution in the center back to the region of small nominal and positive effective densities. In addition, the separation of the bubble can occur unsteadily. In dynamic processes, the formulas (\ref{rpAZ1})-(\ref{rpAZ2}) require corresponding modifications. Another possibility is to describe the transition between the initial and final states in Fig.\ref{f1}d in a non-classical way, as a result of a quantum tunnel transition, as was done by Haggard and Rovelli in \cite{1407.0989}.

To understand the obtained form of the EOS, we will consider Fig.\ref{f1}i,l,j in more detail. For the first two closed states $ p_r $ is everywhere positive and flare-out conditions are not met. For the third closed state, $ p_r $ becomes negative in the region $ f (\rho_{nom})> 36 $. However, for this state $ \rho + p_r \leq0 $ in a narrow region $ f (\rho_{nom}) = 33-35 $, which does not intersect with the previous condition. Therefore, the wormhole for this state is also closed. For the first time, the possibility of negative $ p_r $ at $ f (\rho_{nom})> 34 $ is realized on the black line, and $ p_r $ in it turns out to be negative with the necessary margin corresponding to the throat radius. Also, $ \rho + p_r \leq0 $ for $ f (\rho_{nom}) = 33-35 $ and $ p_t \geq0 $ for $ f (\rho_{nom}) \leq35 $, resulting in bifurcation at $ f (\rho_{nom}) \sim35 $. Further, for open states, the opportunity window expands and $ p_r $ in it reaches the corresponding negative threshold. At the same time, $ p_t $ also goes into the negative area, but later. Thus, there is a region in which the necessary and sufficient conditions for opening a wormhole are satisfied.

To summarize this section, we investigated the quasistatic process of opening a wormhole in the NRDM model, making a corresponding modification of the solution in the region of activation of QG effects. We calculated the corresponding EOS and made sure that it is similar to the previously discussed QG deformed EOS in that at high nominal densities, the EOS components necessarily become negative. In the model we are considering, there is a particular sequence of changing the signs, first the density becomes negative, then the radial pressure, then the tangential one. In this case, a window of opportunities opens for fulfilling the flare-out conditions necessary and sufficient to open the wormhole. An unexpected effect for us was the separation of the bubble of space from the wormhole, which also turned out to be necessary in the considered scheme.

\section{QG wormhole in the center of the Milky Way}\label{sec3}

In this section, we will calculate the opening of the wormhole for the parameters of the central black hole in the Milky Way (MW) galaxy. There is no doubt that this can be done in principle, but we want to control the behavior of all quantities included in the model for such realistic choice of parameters. The calculation result is given in Table~\ref{tab2}. It appears that the behavior of metric profiles and related functions is more sharp than for the model example considered above. The calculation is sensitive to the accuracy of setting Bézier parameters and requires a large number of significant digits to be reproducible. To visualize the result, it is necessary to modify the scaling function by introducing the coefficient $ f (c_f h) $, whose value is chosen so as to provide visualization of both large and small values on one graph. The resulting EOS graphs are structurally similar to those obtained in the above model example. The solution has a bifurcation, after which a window of possibilities opens for fulfilling flare-out conditions leading to the opening of a wormhole. There is also a separation of the bubble of space, whose sizes, as well as the resulting radius of the wormhole throat, are controlled by the model parameters.

As noted in \cite{1701.01569}, the RDM model can be used to describe the distribution of dark matter in galaxies and the associated rotation curves. As a first approximation, the model reproduces the flat rotation curves determined by the parameter $ \epsilon = (v / c)^2 $, which for MW speeds $ v \sim200$~km/s gives $ \epsilon = 4 \cdot10^{-7 } $. A more detailed examination \cite{1903.09972} also allows to describe the deviations of the rotation curves from the flat shape. The $ r_s $ parameter determines the gravitational radius of the central black hole. It was noted in \cite{1812.11801} that, when setting the initial data at large distances, in weak fields, a slightly larger value of $ r_{s, nom} = 1.32 \cdot10^{10} $m should be used to reproduce the observed value $ r_s = 1.2 \cdot10^{10} $m of Ghez et al. \cite{0808.2870}, in strong fields. In this case, the gravitational radius is defined as the point at which the $ b $-profile reaches its maximum, $ r_2 = r_s $. The local properties of the gravitational field, such as the position of the innermost stable circular orbit (ISCO), are related with such $ r_s $, not with $ r_{s, nom}$.

The numerical integration, described in detail in \cite{1701.01569}, shows that after passing the gravitational radius the $ a $- and $ b $-profiles begin to fall very quickly, and the density $ \rho $ grows rapidly, which represents the effect of {\it supershift} \cite{1701.01569} or {\it mass inflation} \cite{0411062}. Just 1000~km after crossing $ r_s $, which is very small compared to $ r_s $ itself, the density reaches Planck values. We chose the attenuation factor $ N = 10 $ and apply the QG cutoff when the density reaches $ \rho_P / N $. Further, $ a $-profile of the RDM model falls into the abyss, reaching values of the order of $ (-10^6) $, but we modify it by choosing a softer profile with the value $ a = -250 $ in the center. Similarly, softer behavior is chosen for $ z $-profile, with the ability to open and close the wormhole. With the chosen parameters, bifurcation occurs at $ r^* = 22$~km, where a bubble and a wormhole throat of such radius form. Further, the radius of the bubble remains unchanged, and the radius of the wormhole throat grows, reaching $ r_0 = 13543$~km in the final state.

\begin{table}
\begin{center}
\caption{QG wormhole in the center of Milky Way galaxy}\label{tab2}

~

\def\arraystretch{1.1}
\begin{tabular}{|l|}
\hline
Model parameters: $\epsilon=4\cdot10^{-7}$, $r_{s,nom}=1.32\cdot10^{10}$m, \\
$r_2=r_s=1.1990455291886923\cdot10^{10}$m
\\ \hline 
QG cutoff: $N=10$, $\rho_P/N=3.82807\cdot10^{68}$m$^{-2}$, \\
$r_{QG}=r_s-1.0003513617763519\cdot10^6$m, $y_{QG}=23.207293345446693$, \\
$a_{QG}=-222.2887073354903$, $z_{QG}=192.8253039716802$, \\ 
$\{y_i,a_i\}=\{\{23.207293345446693, -222.2887073354903\}$,\\
$\{23.207283345446694,-247.28371742005763\}, \{1, -250\}, \{0, -250\}\}$
\\ \hline
Closed state: $\{y_i,z_i\}=\{\{23.207293345446693, 192.8253039716802\}$,\\
$\{23.207193345446694, 442.77560481735327\}, \{15, 300\}, \{10, 200\}, \{5, 100\}$,\\
$\{3, 0.48121182505960347\}, \{0, 0.48121182505960347\}\}$, \\
redshift factor in the center $a(0)=-250$
\\ \hline
Open state: $\{y_i,z_i\}=\{\{23.207293345446693, 192.8253039716802\}$, \\
$\{23.207193345446694, 442.77560481735327\}, \{21.237780648029343, -0.2\}$,\\ 
$\{16.237780648029343, -0.2\}, \{11.237780648029343, -0.2\}$,\\ 
$\{3, 0.48121182505960347\}, \{0, 0.48121182505960347\}\}$, \\
redshift factor in the throat: $a(r_0)=-241.7284225293921$,\\
radius of the throat: $r_0=1.3543177581795398\cdot10^7$m,\\
radius of the bubble: $r^*=22026.465749406787$m
\\ \hline
\end{tabular}

\end{center}
\end{table}

\section{Discussion}\label{sec4}

\begin{figure}
\begin{center}
\includegraphics[width=\textwidth]{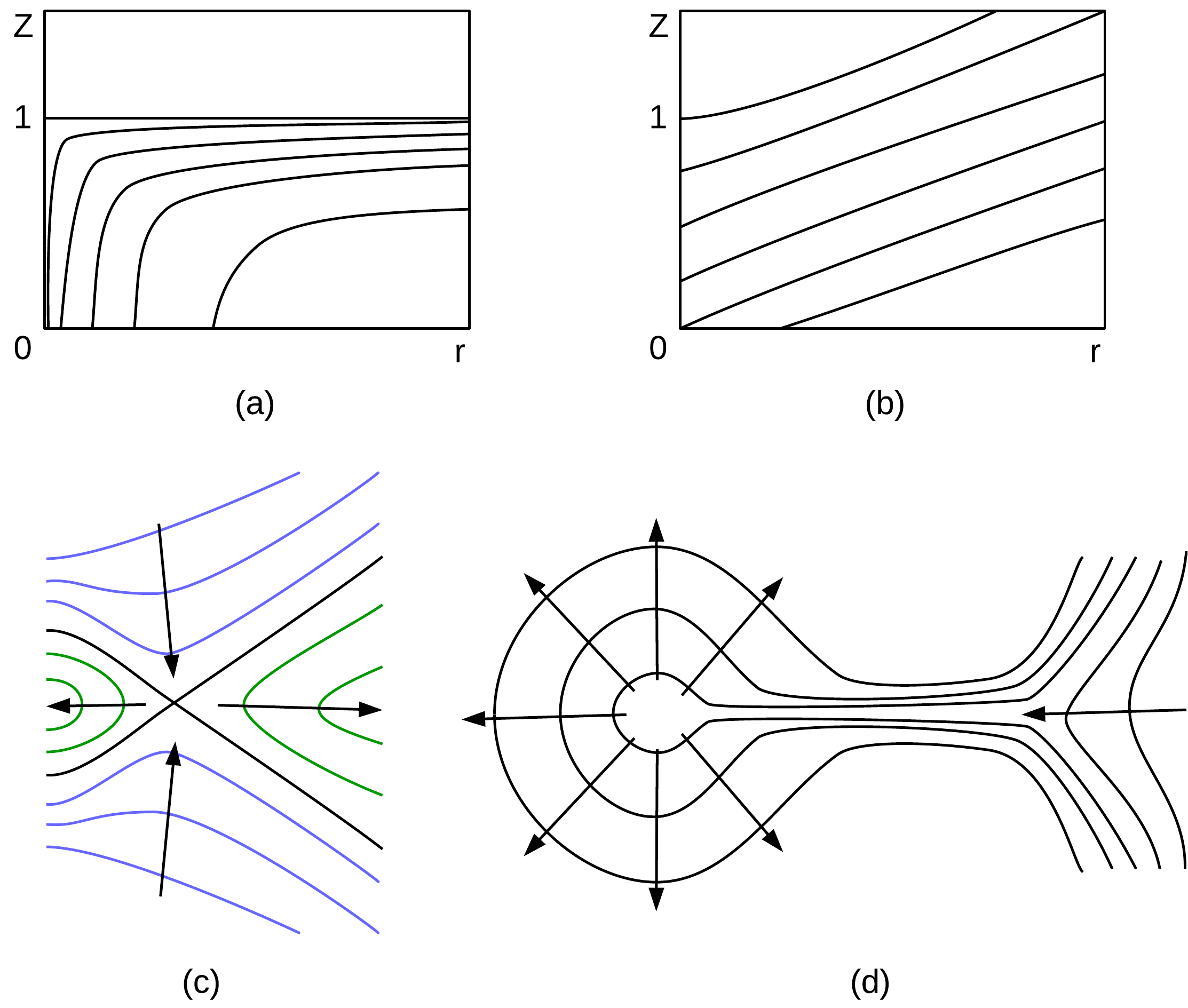}

\end{center}
\caption{Different types of wormhole opening.}\label{f2}
\end{figure}

\paragraph*{Alternative scenarios.}
Fig.\ref{f2}a shows an alternative scenario with an opening a wormhole at $ r_0 = 0 $ based on a family of solutions
\begin{eqnarray}
&&A>0,\ Z=1-2M/r,
\end{eqnarray}
with constant $ M \geq0 $. Solutions of this kind were described in \cite{Visser1996} Chap.13.4.2, for $ M> 0 $ they have the throat in $ r_0 = 2M $. These solutions also have zero energy density, due to the constancy of the mass function, and are fully supported by negative radial pressure. This family can be considered as opening a wormhole with a closed state for $ M = 0 $ and an open state for arbitrarily small $ M> 0 $, that is, arbitrarily small $ r_0> 0 $. Fig.\ref{f2}a shows that when the wormhole opens, the solution has infinitely high derivatives of $ Z $-profile, in contrast to the above solution with separation of the bubble, in which all profiles are regular.

A softer scenario is shown in Fig.\ref{f2}b. The value of $ Z (0) $ gradually descends from 1 to 0, after that the wormhole opens at $ r_0 = 0 $. Next, the point $ Z (r_0) = 0 $ shifts to the right, corresponding to the increase of the radius of the throat. Note that the finiteness of $ \rho (0) $ requires $ Z (0) = 1 $, and as soon as $ Z (0) $ comes off 1, $ \rho (0) $ becomes infinite. However, there is still $ M (0) = 0 $, and this singularity is softer than, for example, the Schwarzschild one with $ M (0) \neq0 $. This solution has the same structure in the center as
\begin{eqnarray}
&&A=Const,\ Z=Const<1,
\end{eqnarray}
solution known as {\it global monopole} by Barriola and Vilenkin \cite{BarriolaVilenkin1989}, see also a review by Tan et al. \cite{1705.00817} (a special case with zero mass, $ M = 0 $).

\begin{figure}
\begin{center}
\includegraphics[width=\textwidth]{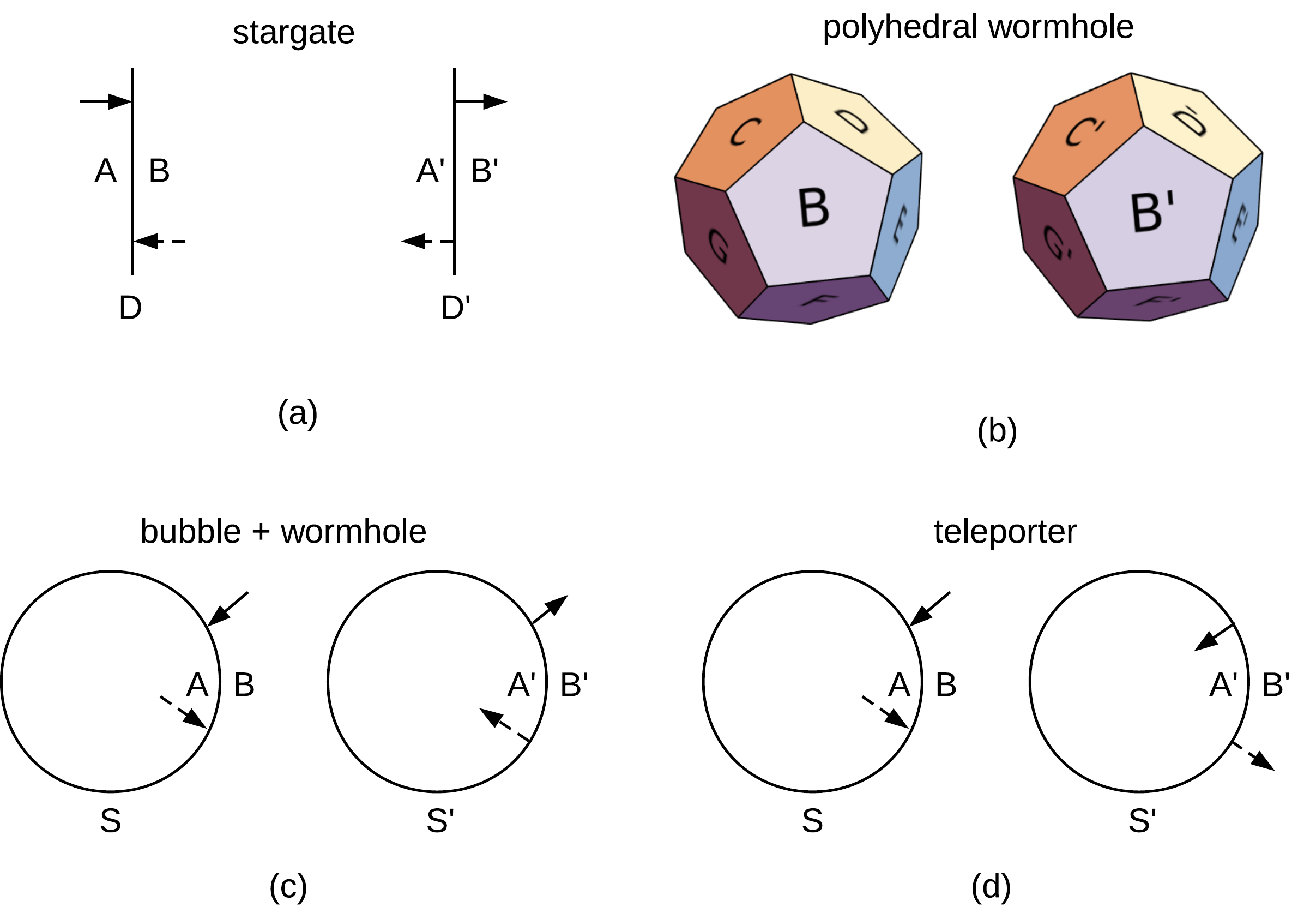}

\end{center}
\caption{Different solutions and their connections.}\label{f3}
\end{figure}

\begin{figure}
\begin{center}
\includegraphics[width=\textwidth]{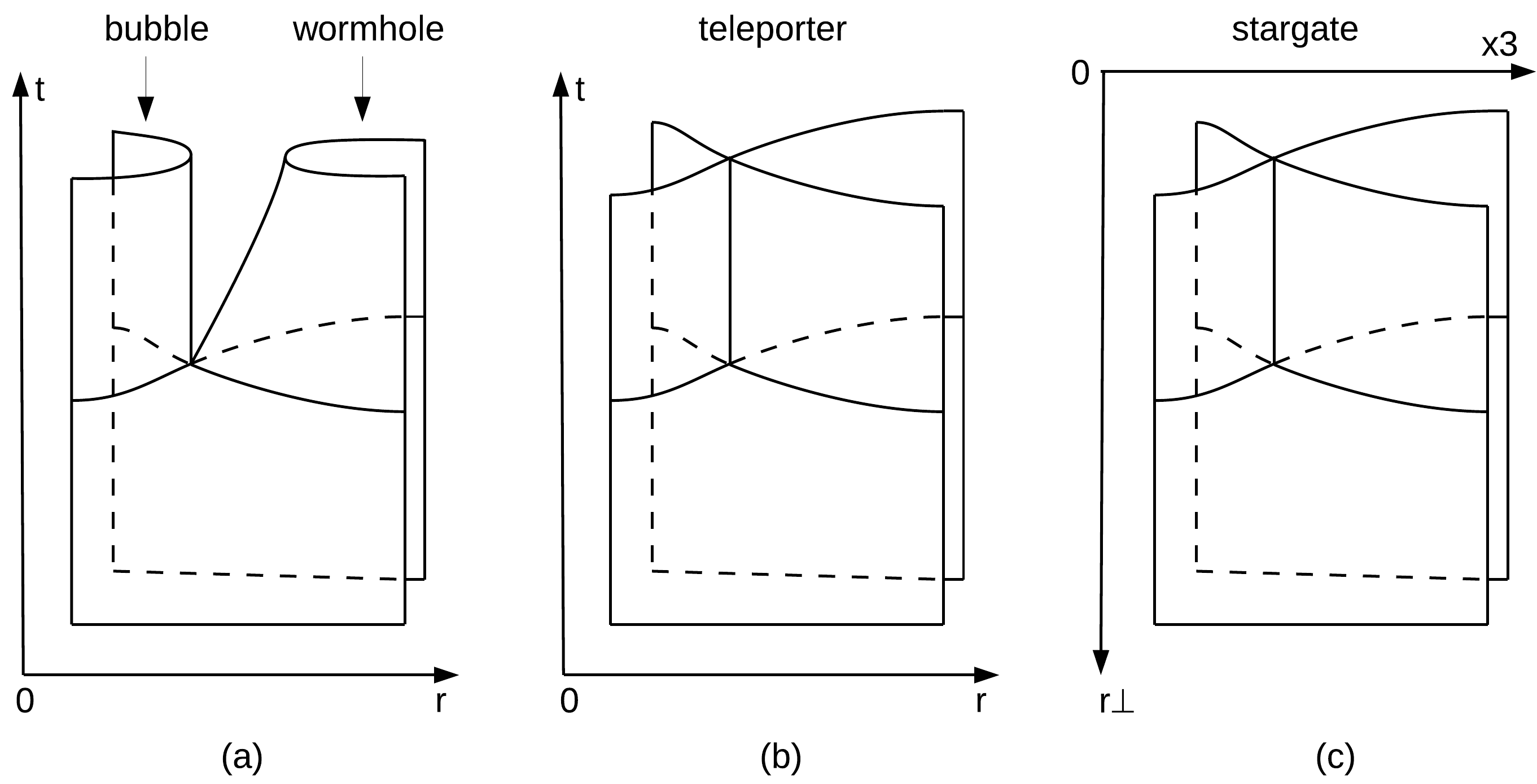}

\end{center}
\caption{Embedding diagrams.}\label{f4}
\end{figure}

Note that any attempt to open a wormhole at $ r_0 = 0 $ and then expand it to $ r_0> 0 $ is associated with singularities, since $ p_r (r_0) = - 1 / (8 \pi r_0^2) $. In this sense, the option we found for opening a wormhole immediately with a finite $ r_0 $ and the formation of a bubble, which we depicted once again in Fig.\ref{f2}c (without preserving some details, such as the invariance of the radius of the bubble), is the softest as it has no singularities in $ a (y) $- and $ z (y) $-profiles and in the associated $ (\rho (y), p_r (y), p_t (y)) $ material parameters.

Fig.\ref{f2}d shows another scenario in which a wormhole and bubble are formed, described in \cite{Visser1996} Chap.9.1.1. In this scenario, the bubble seems to be inflated through a wormhole. Subsequently, the wormhole can be torn, and the bubble is completely separated from the outer space. Similar scenarios were also considered in Battarra et al. \cite{1407.6026} and references therein. They are topologically different from our scenario, whose distinguishing feature is the presence of 2 disjoint patches of $R^3$ in the initial state, while in the final state there are these patches connected by the wormhole and the bubble $S^3$, disconnected from them. Note also that inflation of the bubble through the wormhole goes without changing the topology, but breaking a wormhole requires such a change. If this breaking occurs quasistatically with $ r_0 \to0 $, then it will be accompanied by the singularity in $ p_r $, by the argumentation above.

\paragraph*{Traversability.}
Considering the issue of traversability for wormholes of the considered type, we immediately note that the ultrahigh densities and pressures, as well as the related space curvature inherent to these solutions, are likely to make them non-traversable for human beings, their spaceships and communication signals. Only super creatures capable of withstand such environments, for example, by creating energy shields, can carry out such trips. As for the signals, their known types will certainly be absorbed in the superdense cores of the considered objects. On the other hand, QG wormholes can be opened as a result of natural astrophysical phenomena and can play an important role in the universe, for example, transporting flows of dark matter from one location to the other, disregarding their traversability for human beings.

A more optimistic point of view on this issue is offered by Visser \cite{Visser1996}. According to him, the main obstacle of traversability requiring that the observer should not cross the regions filled with exotic matter, regions of high curvature of spacetime and other dangerous regions is often related to the spherical symmetry. \cite{Visser1996} provides a series of solutions that satisfy all the criteria necessary for traversability. Fig.\ref{f3}a shows a “stargate” solution that has axial, but not spherical symmetry. From two copies of $ R^3 $ space two disks D, D' are cut out, with sides AB and A'B'. Then these sides are glued “crosswise” according to the pattern (AB', BA'), so that the observer crossing the side A of the disk D comes to the side B' of the disk D' and vice versa. Physically equivalent is a disk flipped scheme, (AA', BB'). With these methods of gluing, copies of the space are flat everywhere, except for points on the perimeter of the disks, which are the branch points for the transition between the copies. Therefore, the exotic matter necessary to support such a design is concentrated on the perimeter of the disk, and the calculation of \cite{Visser1996} gives the necessary matter parameters.

This construction can be extended to polyhedra, see Fig.\ref{f3}b. As described in \cite{Visser1996}, identifying the corresponding faces of the polyhedra allows one to obtain the wormholes, in which the exotic matter is concentrated on the edges. Moreover, in the places of gluing on the faces, the space is flat and empty (vacuum), so that the intersection of the faces by the observer is not accompanied by any noticeable effects. With a large number of faces, this solution approaches a spherically symmetric wormhole, but it is granular, leaving vacuum passages interspersed with a concentration of exotic matter on the edges. The spherically symmetric solution is shown in Fig.\ref{f3}c, while gluing BB' creates a wormhole throat, and AA' -- a bubble of space, which we investigated above. For the spherically symmetric solution, the exotic matter is concentrated near the throat or precisely on it if the exotic matter is taken in the form of a thin shell, as described in \cite{Visser1996}.

In addition to the problems of traversability, there is another danger. In the above examples with polyhedra, as well as for spherically symmetric MT wormhole, the passage through the throat is by default accompanied by P-reflection. This is easy to verify, as the angular variables do not change during this transition, and the $ r $-variable changes its direction. An observer emerging from such a wormhole will not only change the left and right sides, all of the internal structure, up to the twist of DNA molecules, will be reversed. Even more serious problems for this scenario arise when considering the properties of P-reflection in particle physics, \cite{Visser1996}. To eliminate these effects, it is better to introduce compensatory P-reflection during the wormhole construction stage.

\paragraph*{Topological Teleporter.} 
Fig.\ref{f3}d shows the gluing method (AB', BA'), dual to that shown in Fig.\ref{f3}c with gluing (AA', BB') for spherical symmetric wormhole solution. It is easy to see that this solution corresponds to instant swapping of two spherical volumes in space, which can be interpreted as an event of teleportation. First, before activating the device, the observer (Captain Kirk) enters via the usual connection BA into the camera of the teleporter S. Then the device is activated, and Kirk exits via the modified connection AB' to the remote location S'. To go back, Kirk takes all the steps in the opposite direction.

This scheme is close to a wormhole, but there are several significant differences. Although the scheme is spherically symmetric, it does not have the throat typical for spherically symmetric wormholes. The space is everywhere flat, with the exception of the SS' spheres at one single moment in time when the teleportation occurs. Accordingly, the matter should be concentrated only at this moment and only on these spheres. This property distinguishes this solution from a thin shell spherically symmetric wormhole, in which the exotic matter is located on the throat sphere for the entire time the wormhole exists. This makes the teleporter solution traversable, for its passage one only needs to avoid being on the SS' spheres at the time of teleportation.

The connection of this solution with the previously considered opening of the wormhole with the bubble is shown in Fig.\ref{f4}. We show embedding diagrams describing these processes. Embedding diagrams are particularly useful for describing dynamic processes where they can replace Bézier curve formulation of static problems when considering B-spline surfaces or other possible representations of two-dimensional surfaces. After specifying the surface, the metric from ambient space is induced onto the surface, the Einstein tensor is calculated, and the matter term necessary to support the solution is determined. Fig.\ref{f4}a shows the process of opening a wormhole with the formation of a bubble and Fig.\ref{f4}b shows a teleportation event. Interestingly, the axisymmetric solution of the stargate type has the same topology, but in different coordinates, see Fig.\ref{f4}c. For the stargate solution, the exotic matter concentrates on the perimeter of the disk all the time, that is, on the cylinder $ S^1 \times R $, while for the teleportation event it concentrates on the sphere $ S^2 $ at one moment in time. These manifolds have the same dimension, but different topology, which is associated with different choices of coordinate systems on the embedding diagrams.

To conclude this section, we will consider another interesting question: can the wormhole and teleporter solutions be used as time machines? The answer is positive and for wormholes it has already been exhaustively investigated in \cite{Visser1996}. Namely, after creating a wormhole, one can control the movement of its end points (mouths) in such a way that the effects of time dilation in special or general theory of relativity will create the time shift necessary for the time machine to function. Entering such a wormhole at the current moment of time, one can exit it at a remote or near point in space, at the past or future moment in time. For the teleporter, when gluing SS' spheres, one can also introduce an arbitrary temporal shift or even T-reflection. Thus, the solution described here, along with wormhole type solutions, allows interesting generalizations and opens up wide opportunities for space engineers of the future.

\section{Conclusion}\label{sec5}
We examined the possibility of opening a wormhole due to QG deformation of EOS. Using the NRDM model as an environment, we showed that a QG wormhole can be opened with a special modification of EOS, which occurs after the nominal classical density exceeds Planck values. In the considered model, under increasing nominal classical density, the components of the energy-momentum tensor become negative in a certain order: first the density $ \rho $ becomes negative, then the radial pressure $ p_r $, then the tangential pressure $ p_t $. In this case, at some point, the flare-out conditions $ p_r=-1/(8\pi r^2)<0 $, $ \rho + p_r <0 $, $ p_t> 0 $, necessary and sufficient to open the wormhole, are satisfied.

In addition, for the quasistatic opening of the wormhole according to the scheme considered in this paper, it is necessary that further $ \rho $ and $ \rho + p_r $ change their sign to positive, while $ p_r $ and $ p_t $ remain negative. In this case, a wormhole is formed with a non-zero radius of the throat, and a bubble separates from the wormhole, encapsulating a region of space within the radius of the throat.

As the calculation shows, these processes proceed both in the model example with a moderate choice of parameters, and when choosing the parameters corresponding to the Milky Way galaxy, while the latter processes proceed in a sharper mode. We also discussed alternatives to the quasistatic opening of a QG wormhole, including the dynamic processes, the quantum tunnel transition, the opening of the wormhole at zero radius with its subsequent expansion, traversability issues and the implementation of various topological connection schemes in this model as well as in related models. In particular, we found a dual connection scheme in which an instant swapping of two volumes in space occurs, which can be interpreted as an event of teleportation. This process is similar to the opening of the wormhol, but differs by the absence of the throat and also has a different matter distribution.

A separate work will be devoted to the study of dynamic solutions and topological changes produced by wormholes and teleporters.

\paragraph*{Acknowledgment.} Many thanks to Lialia Nikitina and Kira Konich for proofreading the paper.

\begin{figure}[h]
\begin{center}
~~\includegraphics[width=0.45\textwidth]{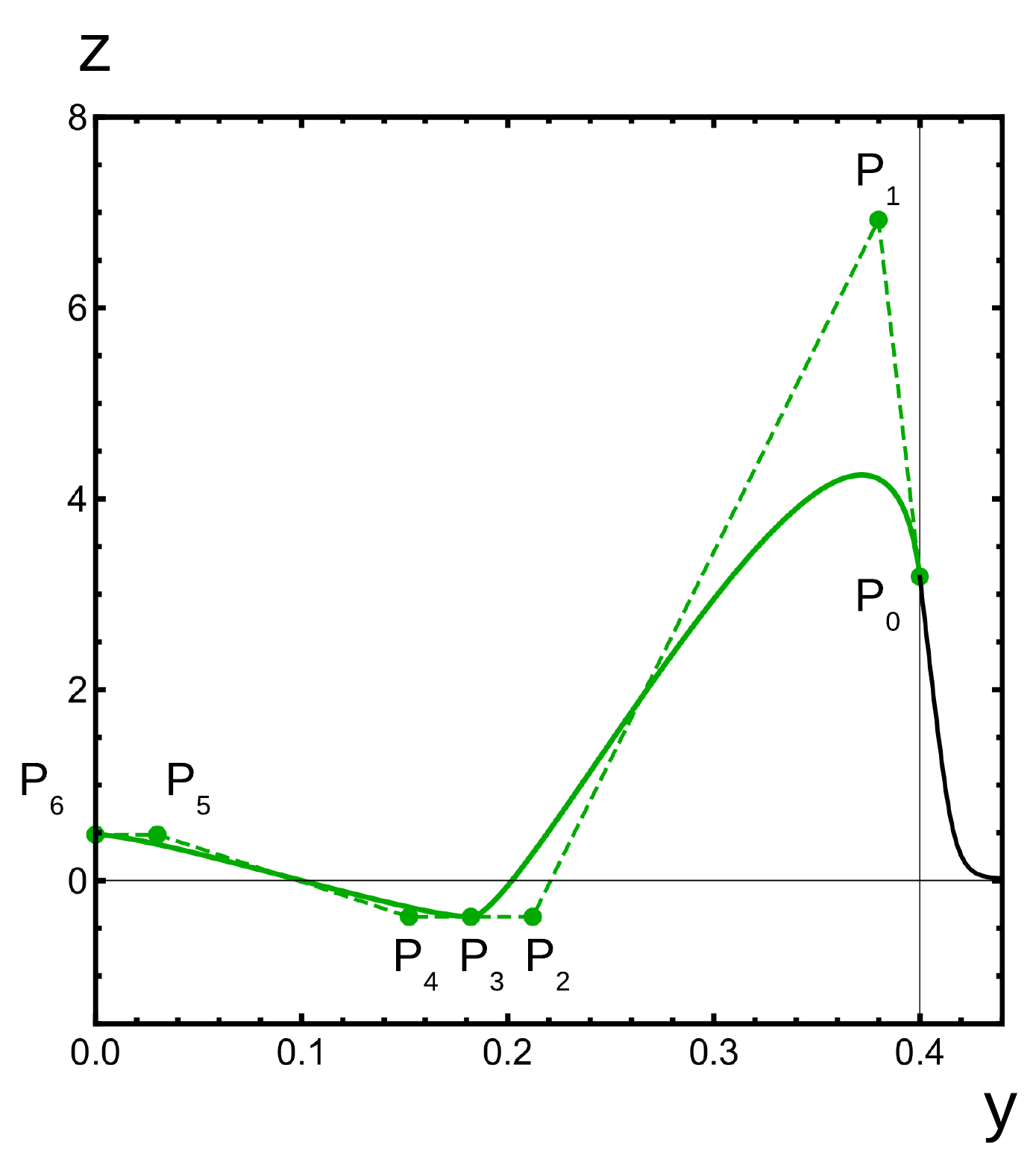}

\end{center}
\caption{Bézier curve setting of metric profile.}\label{f5}
\end{figure}

\section*{Appendix: shaping spacetime with Bézier curves}

For a given set of $ n + 1 $ control points $ \{P_0, P_1, ..., P_n \} $ Bézier curve is specified by the formula \cite{BezierCurve}:
\begin{equation}
X(t)=\sum_{i=0}^n P_i\, B_{i,n}(t),\ B_{i,n}(t)=C^n_i\, t^i(1-t)^{n-i},
\end{equation}
here $ t \in [0,1] $, $ B_{i, n} (t) $ -- Bernstein polynomials, $ C^n_i $ -- binomial coefficients.

The curve starts at $ P_0 $ and ends at $ P_n $, the tangents to the curve at the end points are directed along $ P_1-P_0 $ and $ P_n-P_{n-1} $ respectively. To model complex curves, several Bézier curves are patched together, providing $ C^1 $-smooth stitching between the patches. We use $ n = 3 $ curves to define metric profiles in the coordinates $ a (y) $ and $ z (y) $. An example is shown in Fig.\ref{f5}, the first patch of Bézier curve is set by the points $ \{P_0, P_1, P_2, P_3 \} $, the second is $ \{P_3, P_4, P_5, P_6 \} $. To represent the curves, we use the methods {\tt BezierCurve}, {\tt BezierFunction} and {\tt BernsteinBasis} of {\it Mathematica} system.

\end{document}